\documentclass[intlimits,twoside,a4paper]{article}

\usepackage{amsmath,amssymb}
\usepackage{graphicx}

\usepackage[T2A]{fontenc}
\usepackage[cp1251]{inputenc}
\usepackage{xcolor}

\usepackage[eqsecnum]{cmpj2}

\issue{2015}{18}{1}{13605}
\doinumber{10.5488/CMP.18.13605}

\title[Second-order Barker-Henderson perturbation theory]%
{Second-order Barker-Henderson perturbation theory for the phase behavior of polydisperse
Morse hard-sphere mixture}
\author[T.V. Hvozd, Yu.V. Kalyuzhnyi]{T.V. Hvozd, Yu.V. Kalyuzhnyi}
\address{Institute for Condensed Matter Physics of the National Academy of Sciences of Ukraine, \\ 1 Svientsitskii St., 79011 Lviv, Ukraine
}
\begin{document}

\maketitle

\begin{abstract}

We propose an extension of the second-order Barker-Henderson perturbation theory for polydisperse
hard-sphere multi-Morse mixture. To verify the accuracy of the theory, we compare its predictions
for the limiting case of monodisperse system, with predictions of the very accurate reference
hypernetted chain approximation. The theory is used to describe the liquid--gas phase behavior
of the mixture with different type and different degree of polydispersity. In addition to the
regular liquid--gas critical point, we observe the appearance of the second critical point induced
by polydispersity. With polydispersity increase, the two critical points merge and finally disappear.
The corresponding cloud and shadow curves are represented by the closed curves with `liquid' and
`gas' branches of the cloud curve almost coinciding for higher values of polydispersity. With a further
increase of polydispersity, the cloud and shadow curves shrink and finally disappear.
Our results are in agreement with the results of the previous studies carried out on
the qualitative van der Waals level of description.
\pacs 64.75.-g, 82.70.Dd
\end{abstract}

\section{Introduction}

A vast majority of industrially important colloidal and polymeric materials are intrinsically
polydisperse, i.e., each particle in the system is unique in size, charge, shape, chain length, etc.
This feature of colloidal and polymeric systems has a profound effect on their phase behavior
and may cause the appearance of new phases and new phase transitions.
In addition, the phenomena associated with the phase behavior of polydisperse systems, such as
fractionation, are also of technological relevance.
Usually, the theoretical methods used to study polydisperse systems treat them as a mixture of an
infinite number of components, each characterized by a continuous variable $\xi$ distributed
according to a certain distribution function $f(\xi)$, e.g., for a polydisperse hard-sphere fluid, the hard-sphere size $\sigma$ is usually used
as such a variable, i.e., $\xi=\sigma$.
The theoretical study of the phase
behavior of such fluids, using the methods of the modern liquid state theory, represents a
nontrivial challenge \cite{Sollich2002}.
The main obstacle for theoretical description arises due to the fact that now one has to deal with an
infinite number of equations for coexisting phases. One of the possibilities to overcome this
obstacle is to resort to the so-called truncatable free energy (TFE) models and combine them with the
possibility of their analytical description. TFE models are approximate schemes, where the
thermodynamic properties can be expressed by a finite number of generalized moments of the
distribution function $f(\xi)$. As a result, the formally infinite number of the phase equilibrium
condition equations can be mapped onto a set of a finite number of nonlinear algebraic equations
for these moments and solved using standard numerical methods.
This route was recently undertaken in a number of studies, i.e., phase behavior of Yukawa and
charged hard-sphere polydisperse mixtures were studied using the analytical solution of the
mean spherical approximation (MSA)
\cite{Kalyuzhnyi2003,Kalyuzhnyi2004,Kalyuzhnyi2005a,Kalyuzhnyi2005b} and high temperature
approximation (HTA) \cite{Kalyuzhnyi2006}. More recently, HTA and dimer thermodynamic perturbation theory
(DTPT) for associating fluids in combination with polymer MSA were used to investigate the phase
behavior of a polydisperse mixture of Yukawa chain molecules \cite{Hlushak2007,Hlushak2008}.
While the approaches based on the analytical solution of the MSA appear to be rather accurate, their
application is restricted to the systems with a factorized version of Yukawa interaction,
i.e., the matrix of the coefficients describing the strength of interaction is factorized into the
product of two vectors. On the other hand, HTA based descriptions, being less accurate,
are much more flexible and can be applied to a much larger variety of the potential models
\cite{Kalyuzhnyi2006}.

Attempting to improve the accuracy of the HTA based approaches, we present an
extension and application of the second-order Barker-Henderson perturbation theory for the
phase behavior description of polydisperse multi-Morse hard-sphere mixture. The paper is organized
as follows: In section~\ref{sec:2} we introduce the model and in section~\ref{sec:3} we present a corresponding extension of
the BH2 theory. Our numerical results for the phase behavior of the one-Morse version of the model
are presented and discussed in section~\ref{sec:4}, while in section~\ref{sec:5} we collect our conclusions. In addition, we
include an Appendix with explicit analytical expressions for thermodynamical properties of
the multi-Morse hard-sphere polydisperse mixture in question.

\section{The model}
\label{sec:2}

We consider the mixture with interparticle pair potential represented by the generalized multi-Morse hard-sphere potential
\begin{equation}
\label{MHC1}
  U_\textrm{HSM}(\xi,\xi';r)=\left\{\begin{array}{ll}
                       \infty, & \hbox{$r\leqslant \sigma(\xi,\xi')$}, \\
                       %r\leq \sigma(\xi,\xi') \\
                       -\epsilon_0\sum_{{n=1}}^{N_{M}}\sum_{{m=1}}^{M}(-1)^{m-1}A_{nm}(\xi) A_{nm}(\xi')\re^{-z_n[r-\sigma(\xi,\xi')]}, & \hbox{$r>\sigma(\xi,\xi')$},
                      \end{array}
                   \right.
\end{equation}
where $\xi$ is the polydispersity attribute, i.e., a continuous version of the species index,
$\sigma(\xi)$ is the hard-sphere diameter of the particle of species $\xi$, $\sigma(\xi,\xi')=[\sigma
(\xi)+\sigma(\xi')]/2$, $z_n$ and $\epsilon_0$ are the the screening length and the interaction
strength of the Morse potential, respectively.
The form suggested for the multi-Morse potential (\ref{MHC1}) is similar to that used earlier
for the multi-Yukawa potential \cite{Kalyuzhnyi2006}. This form is very flexible
and can be used to model a large variety of realistic
potentials by an appropriate choice of the coefficients $A_{nm}(\xi)$ and $z_n$, e.g., in
reference \cite{Kalyuzhnyi2006} it is used to mimic
a polydisperse Lennard-Jones {mixture}. Here, $N_M$ denotes the number of the Morse potential tails and
$M$ stands for the number of terms in the sum for one Morse tail.
{Note that original Morse potential consists of two terms, i.e., one is attractive and
the other is repulsive. In our hard-sphere Morse potential repulsive term is substituted by the
hard-sphere term.}

The {mixture} is characterized by the
temperature $T$ [or $\beta=(k_\textrm{B}T)^{-1}$, where $k_\textrm{B}$ is the Boltzmann's constant], the total number-density $\rho$, and by the distribution function  $f(\xi)$ [$\int f(\xi)d\xi=1$].

\section{Theory}
\label{sec:3}

\subsection{Barker-Henderson second-order perturbation theory}

To describe thermodynamic properties of polydisperse Morse hard-sphere {mixture}, we use here the
Barker-Henderson second-order perturbation theory. According to this theory, Helmholtz free energy of
the system $A$ can be written as a sum of three terms: free energy of the reference system
($A_\textrm{ref}$) and the two perturbation terms describing the contribution to the free energy due to Morse
potential ($A_{1},A_{2}$):
\begin{equation}
\label{FRE total}
 A=A_\textrm{ref}+A_{1}+A_{2}=A_\textrm{HS}+A_{1}+A_{2}\,.
\end{equation}
Here, $A_\textrm{ref}=A_\textrm{HS}$, where $A_\textrm{HS}$ is the free energy of the hard-sphere fluid. The first-order term is:
\begin{equation}
\label{FRE1}
\frac{\beta A_{1}}{V}=2\pi\beta\int \rd\xi\int \rd\xi'\rho{(\xi)}\rho{(\xi')}\int_{0}^{\infty} \rd rr^2U_\textrm{HSM}(\xi,\xi';r)g_{(\textrm{HS})}(\xi,\xi';r),
\end{equation}
where $g_{(\textrm{HS})}(\xi,\xi';r)$ is the hard-sphere radial distribution function. For the second-order term, we used the macroscopic compressibility approximation (MCA):
\begin{equation}
\label{FRE2}
\frac{\beta A_{2}}{V}=-\pi\beta^{2}\int \rd\xi\int \rd\xi'\rho{(\xi)}\rho{(\xi')}\int_{0}^{\infty} \rd rr^2\left(\frac{\partial\rho}{\partial p}\right)_\textrm{HS}\left[U_\textrm{HSM}(\xi,\xi';r)\right]^2g_{(\textrm{HS})}(\xi,\xi';r),
\end{equation}
where $\left({\partial\rho}/{\partial p}\right)_\textrm{HS}=\kappa^\textrm{HS}$ is the isothermal compressibility of the hard-sphere reference fluid, which is obtained from the Carnahan-Starling equation and given by
\begin{equation}
\label{compr}
\kappa^\textrm{HS}=\frac{(1-\eta)^{4}}{1+4\eta+4\eta^{2}-4\eta^{3}+\eta^{4}}\,,
\end{equation}
where the packing fraction $\eta$ is defined as
$\eta=\frac{\pi}{6}\int \rd\xi\rho(\xi)\sigma^3(\xi)$.
Substituting into (\ref{FRE1}) and (\ref{FRE2}) the expression for the potential (\ref{MHC1}), we have
\begin{eqnarray}
\label{FRE1a}
\frac{\beta A_{1}}{V}&=&-2\pi\beta\epsilon_{0}\int \rd\xi\int \rd\xi'\rho{(\xi)}\rho{(\xi')} \sum_{{n=1}}^{N_{M}}\sum_{{m=1}}^{M}
(-1)^{m-1}A_{nm}(\xi)A_{nm}(\xi')\nonumber\\
&&\times\left[\sigma(\xi,\xi')\widetilde{G}^{(\textrm{HS})}(\xi,\xi';z_n)
-\frac{\partial\widetilde{G}^{(\textrm{HS})}(\xi,\xi';z_n)}{\partial z_{n}}\right],
\end{eqnarray}
\begin{eqnarray}
\label{FRE2a}
\frac{\beta A_{2}}{V}&=&-\pi\beta^{2}\epsilon_{0}^2\kappa^\textrm{HS}\int \rd\xi\int \rd\xi'\rho{(\xi)}\rho{(\xi')}\sum_{{n=1}}^{N_{M}}\sum_{{m=1}}^{M}
\left[A_{nm}(\xi)A_{nm}(\xi')\right]^2\nonumber\\
&&\times\left[\sigma(\xi,\xi')\widetilde{G}^{(\textrm{HS})}(\xi,\xi';2z_n)
-\frac{\partial\widetilde{G}^{(\textrm{HS})}(\xi,\xi';2z_n)}{\partial (2z_{n})}\right],
\end{eqnarray}
where $\widetilde{G}^{(\textrm{HS})}(\xi,\xi';z_n)$ is the Laplace transform of hard-sphere radial distribution function
\begin{equation}
\label{Gtrdef}
\widetilde{G}^{(\textrm{HS})}(\xi,\xi';z_n)=\re^{z_n\sigma(\xi,\xi')}\int_{0}^{\infty} \rd r r \re^{-z_nr}g_{(\textrm{HS})}(\xi,\xi';r).
\end{equation}
Here, we use Percus-Yevick approximation for the hard-sphere radial distribution function,
since the analytical expressions for its Laplace transform is known.  All the rest thermodynamical quantities
can be obtained using the expression for Helmholtz free energy (\ref{FRE total}) and standard thermodynamical
relations, e.g., differentiating $A$ with respect to the density, we get the expression for the chemical potential:
\begin{equation}
\label{chem total}
\beta\mu(\xi)=\frac{\delta}{\delta\rho(\xi)}\bigg(\frac{\beta A}{V}\bigg),
\end{equation}
and the expression for the pressure $P$ of the system can be calculated invoking the following general
relation:
\begin{eqnarray}
\label{pressure}
\beta P=\beta \int \rd\xi\rho(\xi)\mu(\xi)-\frac{\beta A}{V}\,.
\end{eqnarray}
In the above expressions, $A_\textrm{HS}$ and $\mu^{(\textrm{HS})}(\xi)$ are calculated using the corresponding Mansoori et al. expressions \cite{Mansoori}.

Within the framework of the BH2 approach, the model in question belongs to the class of  `truncatable
free energy models', i.e., the models possessing thermodynamical properties (Helmholtz free energy,
chemical potential, pressure) defined by a finite number of  generalized moments. In this
study, we have the following moments:
\begin{equation}
\label{Mlpol}
m_l=\int \rd\xi\rho(\xi)m_l(\xi)f(\xi), \qquad   m_l(\xi)=\sigma^l,
\end{equation}
\begin{align}
\label{Mlnpol}
& m_l^{(n)}=\int \rd\xi \rho(\xi)m_l^{(n)}(\xi)f(\xi), &  & m_l^{(n)}(\xi)=\sigma^l\varphi(z_n,\sigma),&
& \varphi(z_n,\sigma)=\frac{1}{z_n^2}\left(1-z_n\sigma-\re^{-z_n\sigma}\right),\\
%\end{align}
%\begin{equation}
\label{Mln2pol}
& \widetilde{m}_l^{(n)}=\int \rd\xi\rho(\xi)\widetilde{m}_l^{(n)}(\xi)f(\xi), &    & \widetilde{m}_l^{(n)}(\xi)=\sigma^l\varphi(2z_n,\sigma), & &
\varphi(2z_n,\sigma)=\frac{1}{(2z_n)^2}\left(1-2z_n\sigma-\re^{-2z_n\sigma}\right),
\end{align}
\begin{align}
\label{Mln_1pol}
&\acute{m}_l^{(n)}=\int \rd\xi\rho(\xi)\acute{m}_l^{(n)}(\xi)f(\xi),& & \acute{m}_l^{(n)}(\xi)=\sigma_k^l\varphi_1(z_n,\sigma),& &
\varphi_1(z_n,\sigma)=\re^{-z_n\sigma},\\
%\end{equation}
%\begin{equation}
\label{Mln2_1pol}
&\widetilde{\acute{m}}_l^{(n)}=\int \rd\xi\rho(\xi)\widetilde{\acute{m}}_l^{(n)}(\xi)f(\xi),& & \widetilde{\acute{m}}_l^{(n)}(\xi)=\sigma_k^l\varphi_1(2z_n,\sigma), & &
\varphi_1(2z_n,\sigma)=\re^{-2z_n\sigma},
\end{align}
\begin{align}
\label{Mlnmpol}
&m_l^{(nm)}=\int \rd\xi \rho(\xi)m_l^{(nm)}(\xi)f(\xi), & & m_l^{(nm)}(\xi)=\sigma^lA^{(nm)},\\
%\end{equation}
%\begin{equation}
\label{Mlnm2pol}
& \widetilde{m}_l^{(nm)}=\int \rd\xi\rho(\xi)\widetilde{m}_l^{(nm)}(\xi)f(\xi), & & \widetilde{m}_l^{(nm)}(\xi)=\sigma^l(A^{(nm)})^2.
\end{align}

Closed form analytical expressions for thermodynamical properties (Helmholtz free energy, chemical potential, pressure) in terms of the generalized moments (\ref{Mlpol})--(\ref{Mlnm2pol}) are presented in the Appendix.

\subsection{Phase equilibrium conditions}

The main obstacle in theoretical studies of the phase behavior in polydisperse systems arises due to the fact that one has to deal with an infinite number of equations for coexisting phases. However, for the present `truncatable  free energy model',
these equations can be written as a finite number of equations for the corresponding generalized moments of the distribution function $f(\xi)$ \cite{Bellier1}.

We assume that at a certain temperature $T$, the system, which is characterized by the parent density $\rho^{(0)}$ and parent-phase distribution function $f^{(0)}(\xi)$, separates into $q$ daughter phases with the densities $\rho^{(1)}$, $\rho^{(2)}$, {\ldots}, $\rho^{(q)}$, and $q$ daughter distributions $f^{(1)}(\xi)$, $f^{(2)}(\xi)$, {\ldots}, $f^{(q)}(\xi)$.
All phase equilibrium conditions of the polydisperse system can be simply obtained by generalizing from the multicomponent case via the prescription  $\rho_{i}=\rightarrow \rho f(\xi)\rd\xi$. Due to this substitution, thermodynamic properties (Helmholtz free energy, chemical potential, pressure) become functionals of the distribution function $f(\xi)$. Thermodynamic conditions of phase equilibrium imply the equality of the pressures,
\begin{eqnarray}
\label{equalpressure}
P^{(1)}(T,[f^{(1)}(\xi)])=P^{(2)}(T,[f^{(2)}(\xi)])= {\ldots} =P^{(q)}(T,[f^{(q)}(\xi)]),
\end{eqnarray}
and of the chemical potentials for each species $\xi$.
\begin{eqnarray}
\label{equalchem}
\mu^{(1)}(\xi,T,[f^{(1)}(\xi)])=\mu^{(2)}(\xi,T,[f^{(2)}(\xi)])= {\ldots} =\mu^{(q)}(\xi,T,[f^{(q)}(\xi)]).
\end{eqnarray}
The phase separation constrained by the conservation of the total number of particles of each species $\xi$,
\begin{eqnarray}
\label{conspart}
f^{(0)}(\xi)=\sum_{{\alpha=1}}^{q}f^{(\alpha)}(\xi)x^{(\alpha)},
\end{eqnarray}
where $x^{(\alpha)}=N^{(\alpha)}/N^{(0)}$ is the ratio of the total number of particles, $N^{(\alpha)}$, in phase $\alpha$ to the total number of particles in the parent phase $N^{(0)}$. The conservation of the total volume occupied by the parent phase:
\begin{eqnarray}
\label{consvol}
\upsilon_{0}=\sum_{{\alpha=1}}^{q}\upsilon^{(\alpha)}x^{(\alpha)},
\end{eqnarray}
where $\upsilon^{(\alpha)}=1/\rho^{(\alpha)}$, $(\alpha=1,2,3, {\ldots} , q)$. Finally, the normalization of the $f^{(\alpha)}(\xi)$
\begin{eqnarray}
\label{norm}
\int f^{(\alpha)}(\xi)\rd\xi=1
\end{eqnarray}
in equation (\ref{conspart}) implies the conservation of the total number of particles:
\begin{eqnarray}
\label{consparticl}
1=\sum_{{\alpha=1}}^{q}x^{(\alpha)}.
\end{eqnarray}

In this study, we will consider a two-phase fractionation of polydisperse Morse hard-sphere {mixture} ($\alpha=1, 2$). We assume that thermodynamical properties of the model depend on $K$ generalized moments $m_0$, $m_1$, $m_2$, {\ldots}, $m_K$, which are defined as follows:
\begin{eqnarray}
\label{momdef}
m_k=\rho\int m_k(\xi)f(\xi)\rd\xi, \qquad k \neq 0,
\end{eqnarray}
and $m_0=\rho$.
For the case of two-phase equilibrium ($\alpha=1, 2$), the conditions (\ref{equalpressure})--(\ref{consparticl}) lead to the following set of relations:
\begin{eqnarray}
\label{equalpressure1}
P^{(1)}(T,\{m_k^{(1)}\})=P^{(2)}(T,\{m_{k}^{(2)}\}),
\end{eqnarray}
\begin{eqnarray}
\label{norm1}
\int f^{(\alpha)}(\xi)\rd\xi=1,\qquad \text{for}  \qquad \alpha=1 \qquad \text{or}  \qquad \alpha=2,
\end{eqnarray}
and
\begin{eqnarray}
\label{momfin}
m_{k}^{(1)}=m_{0}^{(1)}\int m_{k}^{(1)}(\xi)f^{(0)}(\xi)H(\xi,T,m_{0}^{(2)},\{m^{(1)}\}\{m^{(0)}\})\rd\xi,
\qquad k\neq0 ,
\end{eqnarray}
where
\begin{eqnarray}
\label{Hxi}
H(\xi,T,m_{0}^{(2)},\{m^{(1)}\}\{m^{(0)}\})=
%\nonumber\\
\frac{\left(\rho^{(1)}-\rho^{(2)}\right)
A_{12}(\xi,T,m_{0}^{(2)},\{m^{(1)}\}\{m^{(0)}\})}
{\left(\rho^{(1)}\rho^{(2)}/\rho^{(0)}-\rho^{(2)}\right)
+\left(\rho^{(1)}-\rho^{(1)}\rho^{(2)}/\rho^{(0)}\right)
A_{12}(\xi,T,m_{0}^{(2)},\{m^{(1)}\}\{m^{(0)}\})}\,,
\end{eqnarray}
\begin{eqnarray}
\label{Axi}
A_{12}(\xi,T,\{m^{(1)}\}\{m^{(2)}\})=\frac{\rho^{(2)}}{\rho^{(1)}}
\exp\left[\mu_\textrm{ex}^{(2)}(\xi,T,\{m^{(2)}\})-\mu_\textrm{ex}^{(1)}(\xi,T,\{m^{(1)}\})\right]
\end{eqnarray}
and $\mu_\textrm{ex}^{(1)}$ is the value of the chemical potential in phase $1$ in excess of its ideal gas value.

The solution of the set of equations (\ref{equalpressure1})--(\ref{momfin}) for a given temperature $T$, for the density of the parent phase $\rho^{(0)}$, and for the parent species distribution function $f^{(0)}(\xi)$, gives us the coexisting densities $\rho^{(\alpha)}$ of the two daughter phases and the corresponding species distribution functions $f^{(\alpha)}(\xi)$, $\alpha=1,2$. The coexisting densities for different temperatures give us binodals, which are terminated  at a temperature for which the density of one of the phases is equal to the density $\rho^{(0)}$ of the parent phase. These termination points form the cloud and shadow coexisting curves that intersect at the critical point  characterized by the critical temperature $T_\textrm{cr}$ and the critical density
$\rho_\textrm{cr}=\rho^{(1)}=\rho^{(2)}=\rho^{(0)}$.
The cloud and shadow curves can be obtained as a special solution of the general
coexisting problem, when the properties of one phase are
equal to the properties of the parent phase: assuming that the
phase $2$ is the cloud phase, i.e., $\rho^{(2)}=\rho^{(0)}$, and following
the above scheme we will end up with the same set of equations, but with $\rho^{(2)}$ and $f^{(2)}(\xi)$ substituted by $\rho^{(0)}$ and $f^{(0)}(\xi)$, respectively.

\section{Results and discussion}
\label{sec:4}

In this section we present numerical results for the phase behavior of polydisperse Morse
hard-sphere {mixture}. For the species distribution function $f(\xi)$, we have chosen a log-normal
distribution, i.e.,
\begin{equation}
\label{distrF}
 f^{(\textrm{LN})}(\xi)=\frac{I}{\sqrt{2\pi \ln I}}\exp\left[-\frac{\ln^{2}(I^{3/2}\xi)}{2\ln I}\right],
\end{equation}
where $I$ is the polydispersity index.
Log-normal distribution frequently occurs at the colloidal
and polymeric processing \cite{Sollich2002}.
Note that in the monodisperse limit ($I=1$), this
distribution is represented by the Dirac delta-function, $\delta(\xi-1)$.
On the contrary, when $I$ becomes very large ($I\gg1$), the above distribution becomes very
wide, increasing hereby the importance of the particles with a large value of $\xi$.
All calculations were carried out for the one-Morse version of the pair potential (\ref{MHC1}), i.e.,
$N_M=1,\;M=1$, with $z_1=1.8\sigma_0$.
We consider two types of polydispersity of the model: polydispersity only in the
strength (amplitude) of the pair potential $A_{11}(\xi)$ and polydispersity in both, amplitude $A_{11}(\xi)$ and hard-sphere size $\sigma(\xi)$. In the former case, we have chosen
$A_{11}(\xi)=A_0\xi$ and $\sigma(\xi)=\sigma_0$, while in the latter case $A_{11}(\xi)=A_0\xi$ and
$\sigma(\xi)=\sigma_0\xi^{1/4}$. Here, $A_0=1$ and $\sigma_0$ is the hard-sphere size for a monodisperse
version of the model at $I=1$, which is used as a distance unit.
{In what follows the density $\rho$ and temperature $T$ are presented in reduced units, i.e., $\rho^*=\rho\sigma_0^3$ and $T^*=k_\textrm{B}T/\epsilon_0$.}

\begin{figure}[!h]
\centerline{\includegraphics[width=0.5\textwidth]{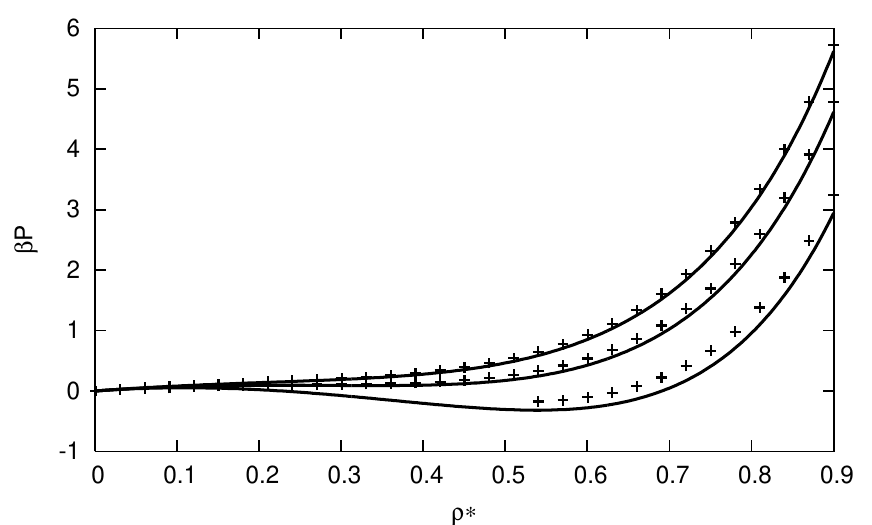}}
\centerline{\includegraphics[width=0.5\textwidth]{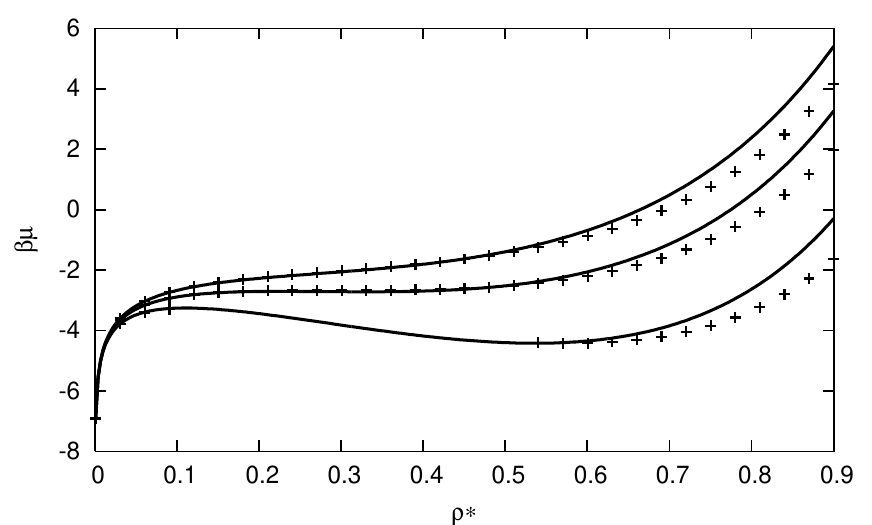}}
\caption{The pressure as a function of the density (the upper panel)
and the chemical potential as a function of the density (the lower panel) along
three isotherms; the upper set of curves corresponds to $T^*=2.5$, the intermediate set
refers to $T^*=2$ and the upper set belongs to $T^*=1.5$. Crosses are RHNC results,
solid lines correspond to the Barker-Henderson second-order results.}
\label{press1}
\end{figure}

As a first step in our numerical study, we perform the calculation of thermodynamical properties
of the monodisperse version of the model, i.e., for $I=1$ using the present version of the
second-order BH
theory and the reference hypernetted chain approximation with the bridge function due to Verlet
\cite{Verlet,Lee}. The latter theory is known for being very accurate in predicting the properties
of simple fluids \cite{Lee}. The comparison of the results of both theories for the pressure
and chemical potential (figure~\ref{press1}) at three different temperatures demonstrate that BH theory is capable
of giving relatively accurate results at lower and intermediate densities. At higher values of the
densities, the predictions of the BH approach are a bit less accurate.

\begin{figure}[!t]
\centerline{\includegraphics[width=0.5\textwidth]{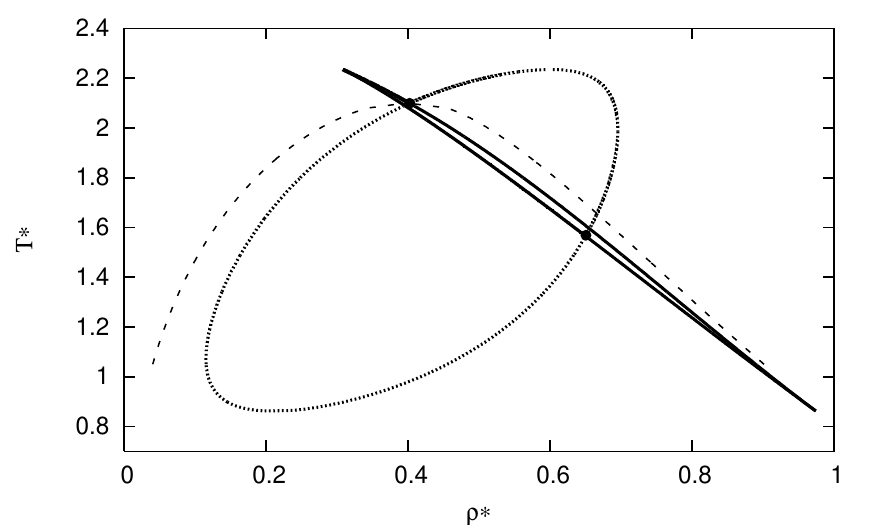}}
\centerline{\includegraphics[width=0.5\textwidth]{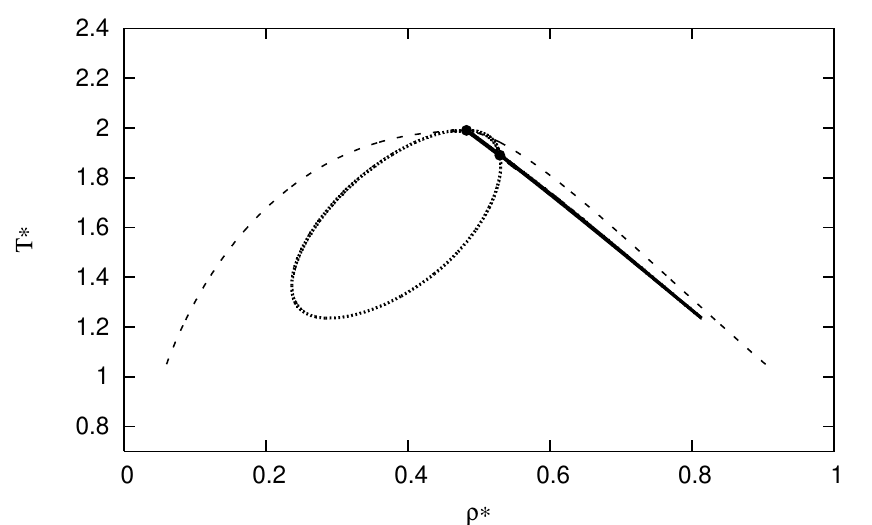}}
\centerline{\includegraphics[width=0.5\textwidth]{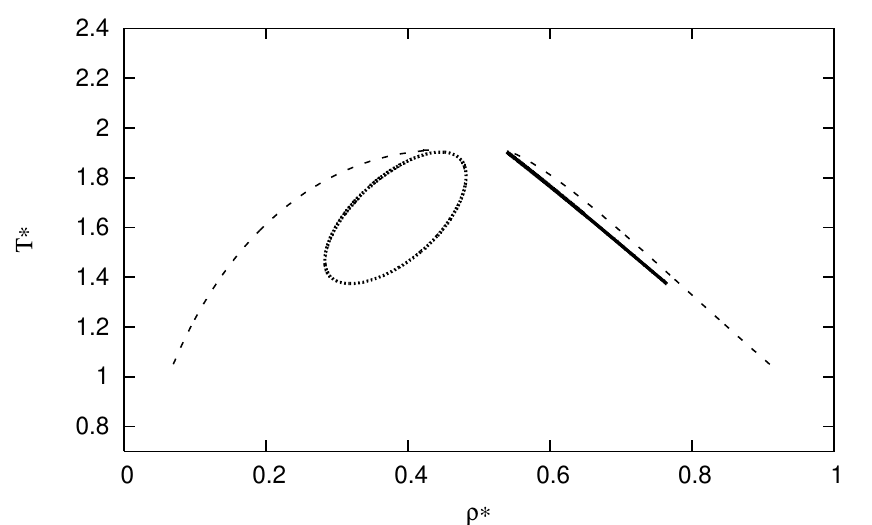}}
\caption{Phase diagrams of the polydisperse Morse hard-sphere {mixture} with amplitude polydispersity only in the $(\rho^*,T^*)$ plane for three different values of the polydispersity index $I$, $I=1.02967$ (the upper panel), $I=1.034$ (the intermediate panel) and $I=1.035$ (the lower panel), obtained using the Barker-Henderson second-order perturbation theory, which includes cloud (solid line) and shadow (dotted line) curves, two critical points and critical binodals (dashed lines). Filled circle denotes the position of the critical points.}
\label{enln}
\end{figure}

\begin{figure}[!t]
\centerline{\includegraphics[width=0.5\textwidth]{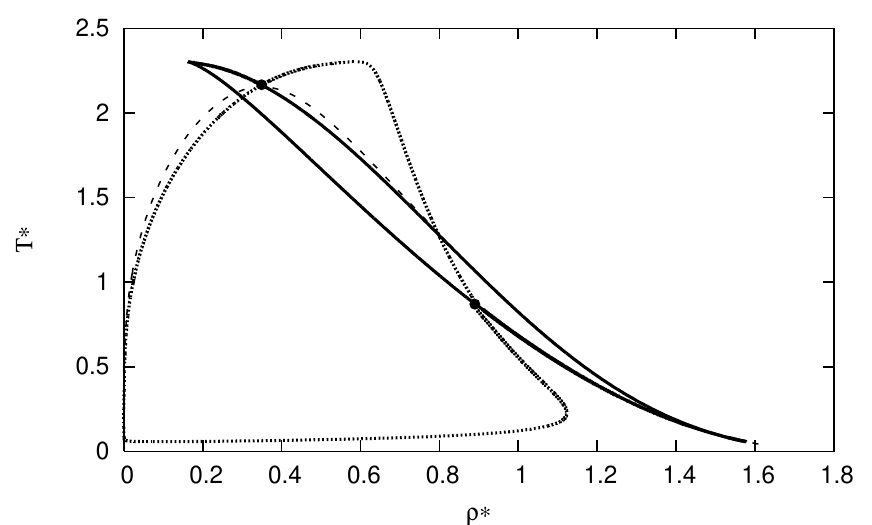}}
\centerline{\includegraphics[width=0.5\textwidth]{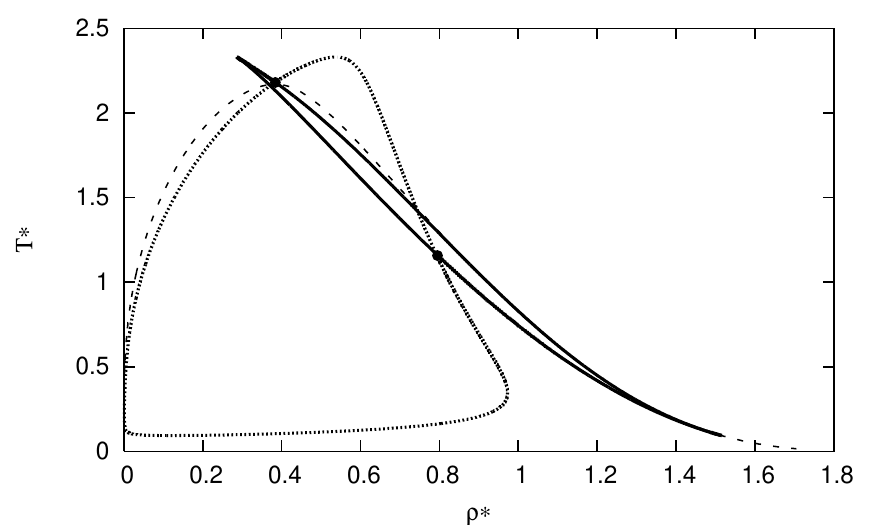}}
\centerline{\includegraphics[width=0.5\textwidth]{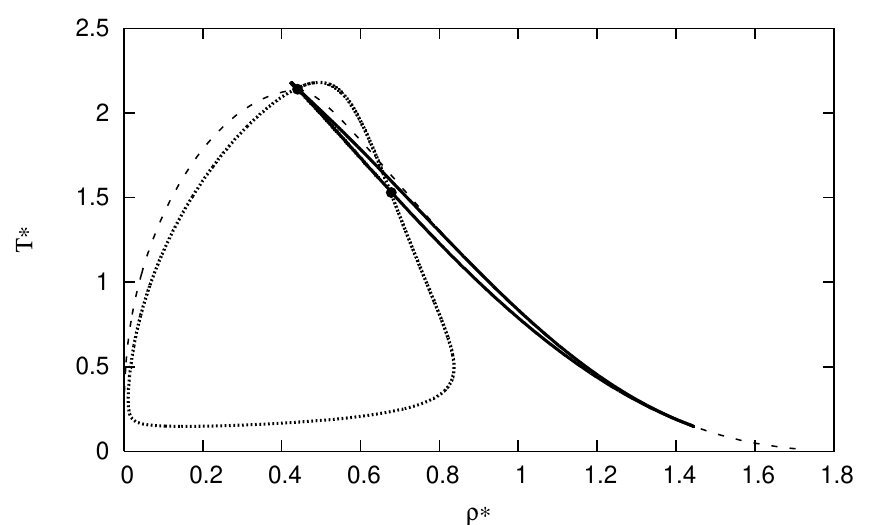}}
\caption{Phase diagrams of the polydisperse Morse hard-sphere {mixture} with size and amplitude polydispersity in the $(\rho^*,T^*)$ plane for three different values of the polydispersity index $I$, $I=1.035$ (the upper panel), $I=1.045$ (the intermediate panel) and $I=1.055$ (the lower panel),  obtained using the Barker-Henderson second-order perturbation theory, which includes cloud (solid line) and shadow (dotted line) curves, two critical points and critical binodals (dashed lines). Filled circle denotes the position of the critical points.}
\label{ensize}
\end{figure}

Next, we perform the calculation of the phase diagram of a polydisperse version of the model at different
values of polydispersity index $I$.
Our results for the model with polydispersity only in the
amplitude of the potential are presented in figure~\ref{enln} and for the model with polydispersity in both
amplitude and hard-sphere size are shown in figure~\ref{ensize}. One can see that polydispersity
has a profound effect on the phase diagram.
When $I=1$, the phase diagram consists of the usual binodal ending in the critical point.
In this case, the cloud and shadow curves coincide, the critical point being located at their maximum.
For a polydisperse system $(I\neq1)$, for each parent phase density $\rho^{(0)}$, there is a
different binodal. Each binodal is truncated at a maximum temperature, with the
corresponding densities, $\rho^{(1)}$ and  $\rho^{(2)}$ lying on the cloud and shadow curves,
respectively. For a critical value of $\rho^{(0)}$, $\rho^{(0)}=\rho_\textrm{cr}$, the corresponding
binodal passes through the intersection of the cloud and shadow curves.
This occurs at $T=T_\textrm{cr}$ and,
since $\rho_\textrm{cr}=\rho^{(1)}=\rho^{(2)}=\rho^{(0)}$, the point $(\rho_\textrm{cr},T_\textrm{cr})$ is a critical
point, where two coexisting phases, $(1)$ and $(2)$, become identical.
We are interested in the phase behavior of the system at relatively large values of
polydispersity. For a small polydispersity, the system has only one critical point
\cite{Kalyuzhnyi2003,Kalyuzhnyi2004,Kalyuzhnyi2005a,Kalyuzhnyi2005b,Kalyuzhnyi2006},
which originates from the regular liquid--gas (LG) critical point of the corresponding monodisperse
version of the system. With the polydispersity increase, there appeares an additional critical
point induced by polydispersity. This effect on the qualitative
van der Waals level of description has been observed by Bellier-Castella et al.
\cite{Bellier1,Bellier2}. The second critical point, which we
denote as polydisperse (P) critical point, is located at larger values of the density and at lower
values of the temperature. It is present for both types of polydispersity studied (figure~\ref{enln} and~\ref{ensize}).
With polydispersity increase, both LG and P  critical points move towards each other and,
for a certain limiting value of polydispersity, they merge. There are no
critical points above this limiting value (figure~\ref{enln}, lower panel). With a further increase of polydispersity and at lower
temperatures, we expect that the two-phase coexistence becomes unstable and there appeares a region
of three-phase coexistence. For relatively high values of polydispersity studied here
we observe a rather unusual shape for the cloud and shadow curves. For both types of polydispersity,
they are represented by closed curves of  elipsoidal- and $\Delta$-like shapes for the shadow
curves seen in figure~\ref{enln} and figure~\ref{ensize}, respectively, and closed curves of the linear shape for the cloud curves (figure~\ref{enln}, \ref{ensize}).
In the latter case, the `liquid' and `gas' branches of the cloud curves almost coincide for the larger
polydispersity  (figure~\ref{enln} and~\ref{ensize}, lower panels). With a further increase of polydispersity, the cloud and
shadow curves shrink and finally disappear.

\begin{figure}[!t]
\centerline{\includegraphics[width=0.5\textwidth]{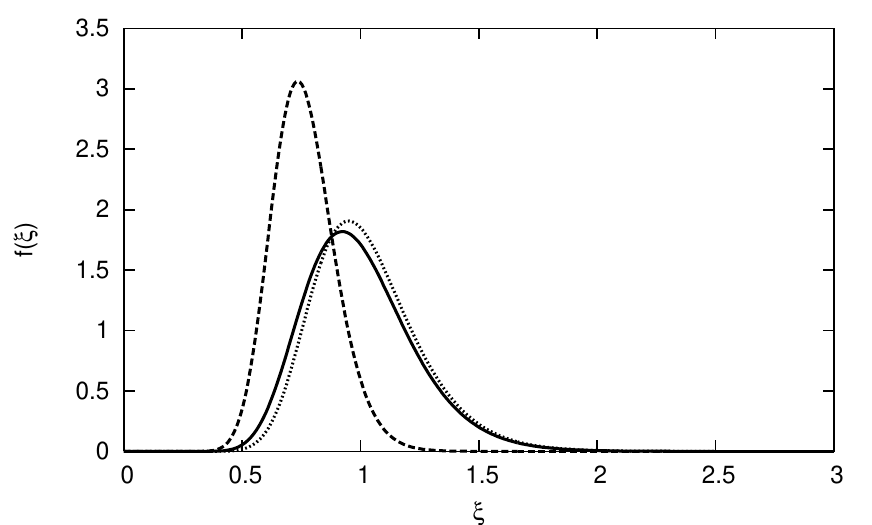}}
\centerline{\includegraphics[width=0.5\textwidth]{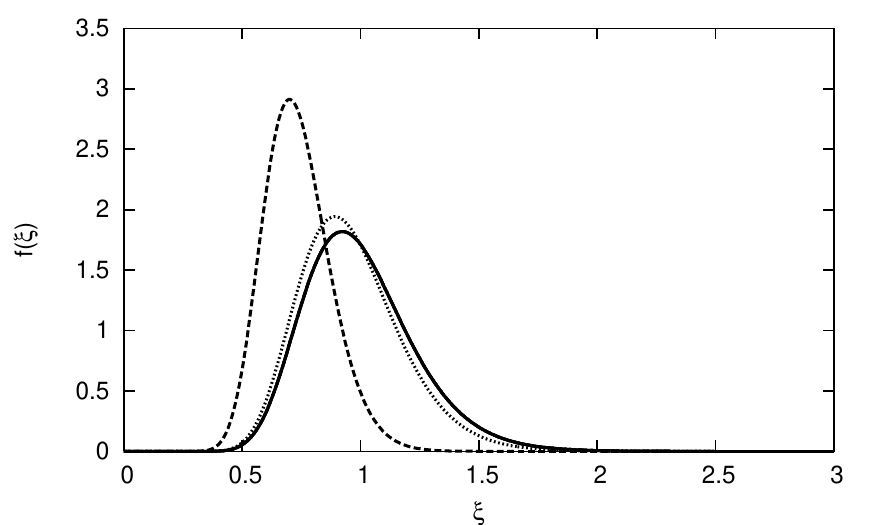}}
\centerline{\includegraphics[width=0.5\textwidth]{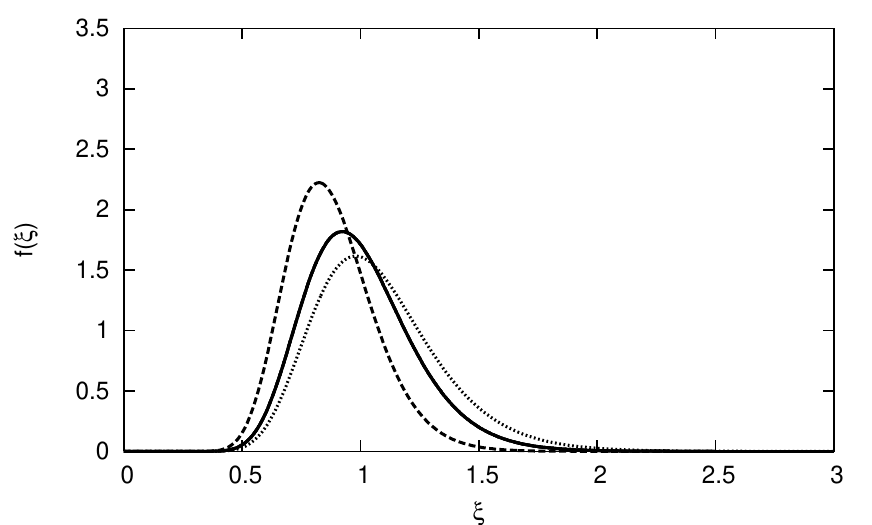}}
\caption{In the upper panel, we display parent $f^{(0)}(\xi)$ (solid line) and daughter $f^{(1)}(\xi)$ (corresponds to gas phase, dashed line) and $f^{(2)}(\xi)$ (corresponds to liquid phase, dotted line) distribution functions for critical binodal at $I=1.055$ and $T^*=1.4$. In the intermediate panel, we show $f^{(1)}(\xi)$ (corresponds to gas shadow phase, dashed line), $f^{(2)}(\xi)=f^{(0)}(\xi)$ (corresponds to liquid and gas cloud phases, solid line) and $f^{(1)}(\xi)$ (corresponds to liquid shadow phase, dotted line) distribution functions at $I=1.055$ and $T^*=1.4$. In the lower panel, we show the same as in intermediate panel at $T^*=1.9$.} \label{distr}
\end{figure}

Finally, in figure~\ref{distr} we display distribution functions of the polydisperse Morse hard-sphere {mixture} with size and amplitude polydispersity at two values of the temperature, which are
higher and lower than the second critical point temperature, i.e., $T^*=1.9$ and $T^*=1.4$,
respectively. We present distribution functions of the coexisting phases on the critical binodal
at $T^*=1.4$ (figure~\ref{distr}, upper panel) and on the coexisting cloud and shadow phases at
$T^*=1.4$ (figure~\ref{distr}, intermediate panel), and $T^*=1.9$ (figure~\ref{distr}, lower panel).
As usual \cite{Kalyuzhnyi2003,Kalyuzhnyi2004,Kalyuzhnyi2005a,Kalyuzhnyi2005b}, on the binodal
the particles with larger values of $\xi$ fractionate to the liquid phase while particles with smaller
values of $\xi$ fractionate into the gas phase. Fractionation of the particles to the
shadow phases for the model at hand depends on the temperature. For the temperatures larger than
the temperature of polydispersity induced critical point, we observe fractionation of the usual
type, i.e., the liquid shadow phase contains particles with  $\xi$ larger than those of the gas cloud
phase while the gas shadow phase contains particles with  $\xi$ smaller than those of the liquid cloud
phase (figure~\ref{distr}, lower panel). Note that distribution functions for the gas and liquid
cloud phases are always the same and equal to the distribution function of the parent phase.
The situation changes for the temperatures lower than
the temperature of the second critical point (figure~\ref{distr}, intermediate panel). In this case,
both liquid and gas shadow phases contain particles with lower values of $\xi$ than those of
the liquid and gas cloud phases. At the same time, the liquid shadow phase has particles of a
larger value of $\xi$ than the gas shadow phase. This behavior of the model is related to the
fact that for temperatures lower than the temperature of the second critical point, both
branches of the shadow curve are located to the left of both branches of the cloud curve, i.e.,
the density of the shadow phases is always lower than the density of the cloud phases.

\section{Conclusions}
\label{sec:5}

In this paper we present an extension of the second-order Barker-Henderson (BH2)
perturbation theory for a polydisperse hard-sphere multi-Morse mixture.
To verify the accuracy of the theory, we compare its predictions
for the limiting case of a monodisperse {system}, with predictions of the very accurate reference
hypernetted chain approximation.
The theory is used to describe
the liquid--gas phase behavior of the mixture with different type and different degree of
polydispersity. In agreement with the previous study \cite{Bellier1,Bellier2},
which was carried out using the qualitative
van der Waals level of description, we observe the appearance of the second critical point induced
by polydispersity. With the polydispersity increase, the two critical points merge and finally disappear,
i.e., for polydispersity larger than a certain threshold value, there are no critical points.
The corresponding cloud and shadow curves are represented by the closed curves with `liquid' and
`gas' branches of the cloud curves that almost coincide with the polydispersity increase. With a further
increase of polydispersity, the cloud and shadow curves shrink and finally disappear.

\appendix

\section*{Appendix}
\renewcommand{\theequation}{A.\arabic{equation}}

Here, we present expressions for thermodynamical properties in terms of the moments
(\ref{Mlpol})--(\ref{Mlnm2pol}). We have:
\begin{eqnarray}
\label{FRE1b}
\frac{\beta A_{1}}{V}\hspace{-2mm}&=&\hspace{-2mm}-2\pi\beta\epsilon_{0}\left\{\frac{Q_{1}^{(nm)}(z_n)}{z_n^2D_0^{(n)}(z_n)}
-\frac{\partial}{\partial z_n}\left[\frac{Q_{0}^{(nm)}(z_n)}{z_n^2D_0^{(n)}(z_n)}\right]\right\}\nonumber\\
\hspace{-2mm}&=&\hspace{-2mm}-\frac{2\pi\beta\epsilon_{0}}{z_n^2D_0^{(n)}(z_n)}\left\{Q_{1}^{(nm)}(z_n)
-\frac{\partial Q_{0}^{(nm)}(z_n)}{\partial z_n}
+Q_{0}^{(nm)}(z_n)\left[\frac{2}{z_n}+\frac{1}{D_0^{(n)}(z_n)}
\frac{\partial D_0^{(n)}(z_n)}{\partial z_n}\right]\right\},
\end{eqnarray}
\begin{eqnarray}
\label{FRE2b}
\hspace{-5mm}\frac{\beta A_{2}}{V}\hspace{-2mm}&=&\hspace{-2mm}-\pi\beta^2\epsilon_{0}^2\kappa^\textrm{HS}
\left\{\frac{\widetilde{Q}_{1}^{(nm)}(2z_n)}{z_n^2\widetilde{D}_0^{(n)}(2z_n)}
-\frac{\partial}{\partial (2z_n)}\left[\frac{\widetilde{Q}_{0}^{(nm)}(2z_n)}{(2z_n)^2\widetilde{D}_0^{(n)}(2z_n)}\right]
\right\}\nonumber\\
\hspace{-2mm}&=&\hspace{-2mm}
-\frac{\pi\beta^2\epsilon_{0}^2\kappa^\textrm{HS}}{4z_n^2\widetilde{D}_0^{(n)}(2z_n)}
\left\{\widetilde{Q}_{1}^{(nm)}(2z_n)-\frac{\partial \widetilde{Q}_{0}^{(nm)}(2z_n)}{\partial (2z_n)}+\widetilde{Q}_{0}^{(nm)}(2z_n)\left[\frac{1}{z_n}
+\frac{1}{\widetilde{D}_0^{(n)}(2z_n)}
\frac{\partial\widetilde{D}_0^{(n)}(2z_n)}{\partial (2z_n)}\right]\right\},
\end{eqnarray}
where
\begin{eqnarray}
\label{D0n}
D_0^{(n)}(z_n)=\Delta^2-\frac{2\pi}{z_n}\left(\Delta+\frac{\pi m_3}{2}\right)\left(m_0^{(n)}+\frac{m_2}{2}\right)-2\pi\left\{\Delta m_1^{(n)}+\frac{\pi}{4}\left[m_2^{(n)}\left(m_2+
+2m_0^{(n)}\right)-2\left(m_1^{(n)}\right)^2\right]\right\},
\end{eqnarray}
\[
\Delta=1-\frac{\pi m_3}{6}\,,
\]
\begin{align}
\label{Q0nm}
Q_0^{(nm)}(z_n)=\Bigg\{&\Delta(m_0^{(nm)})^2+\frac{\pi}{2}\left[\left(m_3+z_nm_2^{(n)}\right)
\left(m_0^{(nm)}\right)^2+\left(\frac{z_n}{2}m_2
+z_nm_0^{(n)}\right)\left(m_1^{(nm)}\right)^2\right]\nonumber\\
&+z_n\left(\Delta-\pi m_1^{(n)}\right)m_1^{(nm)}m_0^{(nm)}\Bigg\},
\end{align}
\begin{align}
\label{Q1nm}
Q_1^{(nm)}(z_n)=\Bigg\{&\Delta m_0^{(nm)}m_1^{(nm)}+\frac{\pi}{2}\left[\left(m_3+z_nm_2^{(n)}\right)m_0^{(nm)}m_1^{(nm)}+\left(\frac{z_n}{2}m_2
+z_nm_0^{(n)}\right) m_1^{(nm)}m_2^{(nm)}\right]
\nonumber\\
&+z_n\left(\Delta-\pi m_1^{(n)}\right)\left(m_1^{(nm)}\right)^2\Bigg\},
\end{align}
\begin{align}
\label{dQ0nmdz}
\frac{\partial Q_0^{(nm)}(z_n)}{\partial z_n}=&\frac{\pi}{2}\left[\left(z_n\frac{\partial m_2^{(n)}}{\partial z_n}+m_2^{(n)}\right)\left(m_0^{(nm)}\right)^2+\left(\frac{m_2}{2}+z_n\frac{\partial m_0^{(n)}}{\partial z_n}+m_0^{(n)}\right)\left(m_1^{(nm)}\right)^2\right]\nonumber\\
&+\left(\Delta-\pi m_1^{(n)}-\pi z_n\frac{\partial m_1^{(n)}}{\partial z_n}\right)m_1^{(nm)}m_0^{(nm)},
\end{align}
\begin{align}
\label{dD0ndz}
\frac{\partial D_0^{(n)}(z_n)}{\partial z_n}=&\frac{2\pi}{z_n^2}\left(\Delta+\frac{\pi m_3}{2}\right)\left(m_0^{(n)}+\frac{m_2}{2}\right)-\frac{2\pi}{z_n}\left(\Delta+\frac{\pi m_3}{2}\right)\frac{\partial m_0^{(n)}}{\partial z_n}\nonumber\\
&-2\pi\left\{\Delta\frac{\partial m_1^{(n)}}{\partial z_n}+\frac{\pi}{4}\left[\frac{\partial m_2^{(n)}}{\partial z_n}\left(m_2+2m_0^{(n)}\right)+2m_2^{(n)}\frac{\partial m_0^{(n)}}{\partial z_n}-
4m_1^{(n)}\frac{\partial m_1^{(n)}}{\partial z_n}\right]\right\},
\end{align}
\begin{eqnarray}
\label{expr1}
m_l^{(n)}=\frac{1}{z_n^2}\left(m_l-z_nm_{l+1}-\acute{m}_l^{(n)}\right),\qquad
m_l^{(n)}(\xi)=\frac{1}{z_n^2}\left[m_l(\xi)-z_nm_{l+1}(\xi)-\acute{m}_l^{(n)}(\xi)\right],
\end{eqnarray}
\begin{eqnarray}
\label{dexpr1dz}
\frac{\partial m_l^{(n)}}{\partial z_n}= \frac{1}{z_n^3}\left(-2m_l+z_nm_{l+1}+2\acute{m}_l^{(n)}+z_n\acute{m}_{l+1}^{(n)}\right).
\end{eqnarray}
Expressions for
$\widetilde{D}_0^{(n)}(2z_n)$, $\widetilde{Q}_0^{(nm)}(2z_n)$, $\widetilde{Q}_1^{(nm)}(2z_n)$, ${\partial \widetilde{Q}_0^{(nm)}(2z_n)}/{\partial (2z_n)}$, ${\partial\widetilde{D}_0^{(n)}(2z_n)}/{\partial (2z_n)}$ in the second-order term of Helmholtz free energy (\ref{FRE2b}) is obtained replacing $z_n$, $m_l^{(n)}$, $\acute{m}_l^{(n)}$, $m_l^{(nm)}$ by $2z_n$, $\widetilde{m}_l^{(n)}$, $\widetilde{\acute{m}}{}_l^{(n)}$, $\widetilde{m}_l^{(nm)}$, respectively, in expressions
(\ref{D0n})--(\ref{dexpr1dz}). Differentiating $\frac{\beta A_1}{V}$ with respect to the density we get the expression for the chemical potential $\beta\mu_1(\xi)$:
\begin{align}
\label{chem1b}
\beta \mu_1(\xi)=&-\frac{2\pi\beta\epsilon_{0}}{z_n^2D_0^{(n)}(z_n)}
\Bigg\{\frac{\delta Q_{1}^{(nm)}(z_n)}{\delta\rho(\xi)}-\frac{\delta}{\delta\rho(\xi)}\left(\frac{\partial Q_{0}^{(nm)}(z_n)}{\partial z_n}\right)+\Bigg(\frac{\delta Q_{0}^{(nm)}(z_n)}{\delta\rho(\xi)}-\frac{Q_{0}^{(nm)}(z_n)}{D_0^{(n)}(z_n)}\nonumber\\
&\times\frac{\delta D_0^{(n)}(z_n)}{\delta \rho(\xi)}\Bigg)\Bigg(\frac{2}{z_n}
+\frac{1}{D_0^{(n)}(z_n)}\frac{\partial D_0^{(n)}(z_n)}{\partial z_n}\Bigg) +\frac{Q_{0}^{(nm)}(z_n)}{D_0^{(n)}(z_n)}\Bigg[\frac{\delta}{\delta\rho(\xi)}
\Bigg(\frac{\partial D_0^{(n)}(z_n)}{\partial z_n}\Bigg) -\frac{1}{D_0^{(n)}(z_n)}\nonumber\\
&\times\frac{\delta D_0^{(n)}(z_n)}{\delta \rho(\xi)}\frac{\partial D_0^{(n)}(z_n)}{\partial z_n}\Bigg]-\frac{Q_{1}^{(nm)}(z_n)}{D_0^{(n)}(z_n)}\frac{\delta D_0^{(n)}(z_n)}{\delta \rho(\xi)}+\frac{1}{D_0^{(n)}(z_n)}\frac{\delta D_0^{(n)}(z_n)}{\delta \rho(\xi)}\frac{\partial Q_0^{(nm)}(z_n)}{\partial z_n}\Bigg\},
\end{align}
\begin{align}
\label{chem2b}
\beta \mu_2(\xi)=&-\frac{\pi\beta^2\epsilon_{0}^2}{(2z_n)^2\widetilde{D}_0^{(n)}(2z_n)}\Bigg(
\kappa^\textrm{HS}\Bigg\{\frac{\delta \widetilde{Q}_{1}^{(nm)}(2z_n)}{\delta\rho(\xi)}-\frac{\delta}{\delta\rho(\xi)}\Bigg(\frac{\partial \widetilde{Q}_{0}^{(nm)}(2z_n)}{\partial (2z_n)}\Bigg)+\Bigg(\frac{\delta \widetilde{Q}_{0}^{(nm)}(2z_n)}{\delta\rho(\xi)}\nonumber\\
&-\frac{\widetilde{Q}_{0}^{(nm)}(2z_n)}{\widetilde{D}_0^{(n)}(2z_n)}
\frac{\delta\widetilde{D}_0^{(n)}(2z_n)}{\delta \rho(\xi)}\Bigg)\Bigg(\frac{1}{z_n}
+\frac{1}{\widetilde{D}_0^{(n)}(2z_n)}\frac{\partial\widetilde{D}_0^{(n)}(2z_n)}{\partial (2z_n)}\Bigg)+\frac{\widetilde{Q}_{0}^{(nm)}(2z_n)}{\widetilde{D}_0^{(n)}(2z_n)}\nonumber\\
&\times\Bigg[\frac{\delta}{\delta\rho(\xi)}\Bigg(\frac{\partial\widetilde{D}_0^{(n)}(2z_n)}{\partial (2z_n)}\Bigg)-\frac{1}{\widetilde{D}_0^{(n)}(2z_n)}\frac{\delta\widetilde{D}_0^{(n)}(2z_n)}{\delta \rho(\xi)}\frac{\partial\widetilde{D}_0^{(n)}(2z_n)}{\partial (2z_n)}\Bigg]\nonumber\\
&-\frac{\widetilde{Q}_{1}^{(nm)}(2z_n)}{\widetilde{D}_0^{(n)}(2z_n)}\frac{\delta\widetilde{D}_0^{(n)}(2z_n)}{\delta \rho(\xi)}+\frac{1}{\widetilde{D}_0^{(n)}(2z_n)}\frac{\delta\widetilde{D}_0^{(n)}(2z_n)}{\delta \rho(\xi)}\frac{\partial \widetilde{Q}_0^{(nm)}(2z_n)}{\partial (2z_n)}\Bigg\}\nonumber\\
&+\frac{\delta\kappa^\textrm{HS}}{\delta\rho(\xi)}
\Bigg[\widetilde{Q}_{1}^{(nm)}(2z_n)
-\frac{\partial \widetilde{Q}_{0}^{(nm)}(2z_n)}{\partial (2z_n)}+\widetilde{Q}_{0}^{(nm)}(2z_n)
\Bigg
(\frac{1}{z_n}+\frac{1}{\widetilde{D}_0^{(n)}(2z_n)}\frac{\partial\widetilde{D}_0^{(n)}(2z_n)}{\partial (2z_n)}\Bigg)\Bigg]\Bigg),
\end{align}
where
\begin{align}
\label{dQ0dro}
\frac{\delta Q_0^{(nm)}(z_n)}{\delta \rho(\xi)}=& %\Bigg\{
\pi\left(\frac{m_3(\xi)}{3}+\frac{z_nm_2^{(n)}(\xi)}{2}\right)\left(m_0^{(nm)}\right)^2
+\left(2\Delta+\pi m_3+\pi z_nm_2^{(n)}\right)m_0^{(nm)}m_0^{(nm)}(\xi)\nonumber\\
&+\pi z_n \bigg[\frac{1}{2}\left(\frac{m_2(\xi)}{2}+m_0^{(n)}(\xi)\right)\left(m_1^{(nm)}\right)^2
+\left(\frac{m_2}{2}+m_0^{(n)}\right)
m_1^{(nm)}m_1^{(nm)}(\xi)\nonumber\\
&-\left(\frac{m_3(\xi)}{6}+m_1^{(n)}(\xi)\right)m_0^{(nm)}m_1^{(nm)}
+\left(\frac{\Delta}{\pi}-m_1^{(n)}\right)
\left(m_0^{(nm)}m_1^{(nm)}(\xi)+m_1^{(nm)}m_0^{(nm)}(\xi)\right)\bigg]%\Bigg\}
,
\end{align}
\begin{align}
\label{dQ1dro}
\frac{\delta Q_1^{(nm)}(z_n)}{\delta \rho(\xi)}=& %\Bigg\{
\pi\left(\frac{m_3(\xi)}{3}+\frac{z_nm_2^{(n)}(\xi)}{2}\right)m_0^{(nm)}m_1^{(nm)}+
\left(\Delta+\frac{\pi m_3}{2}+\frac{\pi z_nm_2^{(n)}}{2}\right)\nonumber\\ &\times\left(m_0^{(nm)}m_1^{(nm)}(\xi)+m_1^{(nm)}m_0^{(nm)}(\xi)\right)+\pi z_n \bigg[\frac{1}{2}\left(\frac{m_2(\xi)}{2}+m_0^{(n)}(\xi)\right)m_1^{(nm)}m_2^{(nm)}\nonumber\\
&+\frac{1}{2}\left(\frac{m_2}{2}+m_0^{(n)}\right)\left(m_1^{(nm)}m_2^{(nm)}(\xi)
+m_2^{(nm)}m_1^{(nm)}(\xi)\right)
-\left(\frac{m_3(\xi)}{6}+m_1^{(n)}(\xi)\right)(m_1^{(nm)})^2\nonumber\\
&+\left(\frac{\Delta}{\pi}-m_1^{(n)}\right)2m_1^{(nm)}m_1^{(nm)}(\xi)\bigg]
%\Bigg\}
,
\end{align}
\begin{align}
\label{dD0}
\frac{\delta D_0^{(n)}(z_n)}{\delta\rho(\xi)}=
&2\pi\Bigg\{\frac{1}{3}\pi m_3(\xi)\left[\frac{1}{2}m_1^{(n)}-\frac{1}{z_n}\left(m_0^{(n)}+\frac{1}{2}m_2\right)\right]-
\Delta\left(\frac{1}{6}m_3(\xi)++m_1^{(n)}(\xi)\right)\nonumber\\
&-\left(\frac{1}{2}m_2(\xi)+m_0^{(n)}(\xi)\right)\bigg[\frac{1}{z_n}\left(\Delta+\frac{1}{2}\pi m_3\right)+\frac{1}{2}\pi m_2^{(n)}\bigg]\nonumber\\
&-\frac{1}{4}\pi m_2^{(n)}(\xi)\left(m_2+2m_0^{(n)}\right)+\pi m_1^{(n)}m_1^{(n)}(\xi)\Bigg\},
\end{align}
\begin{align}
\label{dQ0nmdzdro}
\frac{\delta}{\delta\rho(\xi)}&\left(\frac{\partial Q_0^{(nm)}(z_n)}{\partial z_n}\right)=\frac{\pi}{2}\Bigg\{\left[z_n\frac{\delta}{\delta\rho(\xi)}\left(\frac{\partial m_2^{(n)}}{\partial z_n}\right)+m_2^{(n)}(\xi)\right]\left(m_0^{(nm)}\right)^2
+\left(z_n\frac{\partial m_2^{(n)}}{\partial z_n}+m_2^{(n)}\right)2m_0^{(nm)}m_0^{(nm)}(\xi)\nonumber\\
&+\left[\frac{m_2(\xi)}{2}+z_n\frac{\delta}{\delta\rho(\xi)}\left(\frac{\partial m_0^{(n)}}{\partial z_n}\right)+m_0^{(n)}(\xi)\right]\left(m_1^{(nm)}\right)^2+\left(\frac{m_2}{2}
+z_n\frac{\partial m_0^{(n)}}{\partial z_n}+m_0^{(n)}\right)2m_1^{(nm)}m_1^{(nm)}(\xi)\Bigg\}\nonumber\\
&-\left[\frac{\pi m_3(\xi)}{6}+\pi m_1^{(n)}(\xi)+\pi z_n\frac{\delta}{\delta\rho(\xi)}\left(\frac{\partial m_1^{(n)}}{\partial z_n}\right)\right]m_1^{(nm)}m_0^{(nm)}\nonumber\\
&+\left(\Delta-\pi m_1^{(n)}-\pi z_n\frac{\partial m_1^{(n)}}{\partial z_n}\right)\left(m_1^{(nm)}m_0^{(nm)}(\xi)+m_1^{(nm)}(\xi)m_0^{(nm)}\right),
\end{align}
\begin{align}
\label{dD0ndzdro}
\frac{\delta}{\delta\rho(\xi)}&\left(\frac{\partial D_0^{(n)}(z_n)}{\partial z_n}\right)=\frac{2\pi}{z_n^2}\left[\frac{\pi m_3(\xi)}{3}\left(m_0^{(n)}+\frac{m_2}{2}\right)+
\left(\Delta+\frac{\pi m_3}{2}\right)\left(m_0^{(n)}(\xi)+\frac{m_2(\xi)}{2}\right)\right]
\nonumber\\
&-\frac{2\pi}{z_n}\left[\frac{\pi m_3(\xi)}{3}\frac{\partial m_0^{(n)}}{\partial z_n}+
\left(\Delta+\frac{\pi m_3}{2}\right)\frac{\delta}{\delta\rho(\xi)}\left(\frac{\partial m_0^{(n)}}{\partial z_n}\right)\right]-2\pi\Bigg\{-\frac{\pi m_{3}(\xi)}{6}\frac{\partial m_1^{(n)}}{\partial z_n}+\Delta\frac{\delta}{\delta\rho(\xi)}\left(\frac{\partial m_1^{(n)}}{\partial z_n}\right)\nonumber\\
&+\frac{\pi}{4}\Bigg[\frac{\partial m_2^{(n)}}{\partial z_n}\left(m_2(\xi)+2m_0^{(n)}(\xi)\right)
+\left(m_2+2m_0^{(n)}\right)\frac{\delta}{\delta\rho(\xi)}\left(\frac{\partial m_2^{(n)}}{\partial z_n}\right)+2m_2^{(n)}(\xi)\frac{\partial m_0^{(n)}}{\partial z_n}\nonumber\\
&+2m_2^{(n)}\frac{\delta}{\delta\rho(\xi)}\left(\frac{\partial m_0^{(n)}}{\partial z_n}\right)
-4m_1^{(n)}(\xi)\frac{\partial m_1^{(n)}}{\partial z_n}
-4m_1^{(n)}\frac{\delta}{\delta\rho(\xi)}\left(\frac{\partial m_1^{(n)}}{\partial z_n}\right)\Bigg]\Bigg\},
\end{align}
\begin{eqnarray}
\label{dexpr1dzdro}
\frac{\delta}{\delta\rho(\xi)}\left(\frac{\partial m_l^{(n)}}{\partial z_n}\right)= \frac{1}{z_n^3}\left(-2m_l(\xi)+z_nm_{l+1}(\xi)+2\acute{m}_l^{(n)}(\xi)
+z_n\acute{m}_{l+1}^{(n)}(\xi)\right).
\end{eqnarray}
\begin{eqnarray}
\label{dcompr}
\frac{\delta\kappa^\textrm{HS}}{\delta\rho(\xi)}=-\frac{2\pi m_3(\xi)}{3}\left[
\frac{(1-\eta)^{3}}{1+4\eta+4\eta^{2}-4\eta^{3}+\eta^{4}}
+\frac{\left(1+2\eta-3\eta^2+\eta^3\right)(1-\eta)^{4}}
{\left(1+4\eta+4\eta^{2}-4\eta^{3}+\eta^{4}\right)^2}\right],
\end{eqnarray}
Expressions for
\[
\frac{\delta\widetilde{D}_0^{(n)}(2z_n)}{\delta\rho(\xi)}, \quad \frac{\delta\widetilde{Q}_0^{(nm)}(2z_n)}{\delta\rho(\xi)}, \quad \frac{\delta\widetilde{Q}_1^{(nm)}(2z_n)}{\delta\rho(\xi)}, \quad \frac{\delta}{\delta\rho(\xi)}\left[\frac{\partial \widetilde{Q}_0^{(nm)}(2z_n)}{\partial (2z_n)}\right], \quad \frac{\delta}{\delta\rho(\xi)}\left[\frac{\partial\widetilde{D}_0^{(n)}(2z_n)}{\partial (2z_n)}\right]
\]
in the second-order term for chemical potential (\ref{chem2b}) are obtained by replacing $z_n$, $m_l^{(n)}$, $m_l^{(n)}(\xi)$, $\acute{m}_l^{(n)}$, $\acute{m}_l^{(n)}(\xi)$, $m_l^{(nm)}$, $m_l^{(nm)}(\xi)$ by $2z_n$, $\widetilde{m}_l^{(n)}$, $\widetilde{m}_l^{(n)}(\xi)$, $\widetilde{\acute{m}}{}_l^{(n)}$, $\widetilde{\acute{m}}{}_l^{(n)}(\xi)$, $\widetilde{m}_l^{(nm)}$, $\widetilde{m}_l^{(nm)}(\xi)$, respectively, in expressions (\ref{dQ0dro})--(\ref{dexpr1dzdro}).

\ukrainianpart

\title{Застосування термодинамічної теорії збурень другого порядку Баркера-Хендерсона
для дослідження фазової поведінки полідисперсної {суміші} твердих сфер Морзе}
%\author{А.В. Тор\refaddr{label1,label2}, Б.В. Тор\refaddr{label2}}
%\addresses{
%\addr{label1} Університет ім. Орнштейна, Софтленд, 10041 Цельсій, вул. Реін, 1
%\addr{label2} Інститут ім. Церніке, Солідшир, 20451 Фаренгейт, пр. Рівер, 2
%}
%
%% якщо автор є один або автори є з однієї установи:
%
%  \author{1й Автор, 2й Автор, \ldots}
%  \address{Інститут\ldots}
%
%%

\author{Т.В. Гвоздь, Ю.В. Калюжний}
\address{Інститут фізики конденсованих систем НАН України, вул. І. Свєнціцького, 1,  79011 Львів, Україна}

\makeukrtitle

\begin{abstract}
\tolerance=3000%
Запропоновано застосування термодинамічної теорії збурень другого порядку Баркера-Хендерсона
для дослідження полідисперсної суміші твердих сфер Морзе. Для перевірки точності  порівнюються  результати цієї теорії для граничного випадку монодисперсної
{системи} з результатами дуже точного базисного гіперланцюжкового наближення. Теорія використовується для опису фазової поведінки рідина-газ для суміші з різними типами та різними ступенями полідисперсності. Окрім звичайної критичної точки рідина--газ, ми спостерігаємо появу другої критичної точки, яка є зумовлена полідисперсністю. Із збільшенням полідисперсності ці дві критичні точки зливаються і, нарешті, зникають.
Відповідні криві хмари та тіні представлені замкненими кривими з гілками рідина та газ, для кривої хмари вони майже збігаються для вищих значень полідисперсності. При подальшому збільшенні полідисперсності криві хмари та тіні скорочуються і, нарешті,  зникають.
Наші результати узгоджуються з результатами попередніх досліджень, які були проведені на якісному рівні опису ван дер Ваальса.

\keywords термодинамічна теорія збурень, полідисперсність, фазове співіснування, колоїдні системи, потенціал Морзе

\end{abstract}

\end{document}